\documentclass[twocolumn,nopacs,floatfix,amsmath,nofootinbib,amssymb,floatfix]{revtex4}
\usepackage{graphicx,color,dcolumn,booktabs,bm}
\usepackage{longtable,lscape}
\usepackage{txfonts}
\usepackage{overpic}
\usepackage{amssymb}
\usepackage{indentfirst}
\usepackage{feynmf}   
\usepackage{slashed}  
\usepackage{cases}
\usepackage{color}
\usepackage{multirow}
\usepackage{longtable}
\usepackage{ulem}
\usepackage{enumerate}
\usepackage{graphicx,color,dcolumn,booktabs,bm}
\usepackage[colorlinks,
            citecolor=blue,
            anchorcolor=red,
            menucolor=red,
            linkcolor=red,
            filecolor=red,
            runcolor=red,
            urlcolor=blue,
            frenchlinks=red]{hyperref}
\usepackage{epstopdf}
\usepackage{array}
\begin{document}
\title{Predicting a new resonance as charmed-strange baryonic analog of $D_{s0}^*(2317)$}

\author{Si-Qiang Luo$^{1,4}$}\email{luosq15@lzu.edu.cn}
\author{Bing Chen$^{2,3}$}\email{chenbing@ahstu.edu.cn}
\author{Xiang Liu$^{1,3,4}$}\email{xiangliu@lzu.edu.cn}
\author{Takayuki Matsuki$^{5}$}\email{matsuki@tokyo-kasei.ac.jp}
\affiliation
{
$^1$School of Physical Science and Technology, Lanzhou University, Lanzhou 730000, China\\
$^2$School of Electrical and Electronic Engineering, Anhui Science and Technology University, Fengyang 233100, China\\
$^3$Lanzhou Center for Theoretical Physics, Key Laboratory of Theoretical Physics of Gansu Province, and Frontiers Science Center for Rare Isotopes, Lanzhou University, Lanzhou 730000, China\\
$^4$Research Center for Hadron and CSR Physics, Lanzhou University $\&$ Institute of Modern Physics of CAS,
Lanzhou 730000, China\\
$^5$Tokyo Kasei University, 1-18-1 Kaga, Itabashi, Tokyo 173-8602, Japan}

\begin{abstract}
By an unquenched quark model, we predict a charmed-strange baryon state, namely, the $\Omega_{c0}^d(1P,1/2^-)$. Its mass is predicted to be 2945 MeV, which is below the $\Xi_c\bar{K}$ threshold due to the nontrivial coupled-channel effect. So the $\Omega_{c0}^d(1P,1/2^-)$ state could be regraded as the analog of the charmed-strange meson $D_{s0}^*(2317)$. It is a good opportunity for the running Belle II experiment to search for this state in the $\Omega_c^{(*)}\gamma$ mass spectrum experiment in the future.
\end{abstract}

\maketitle

\section{Introduction}\label{introduction}

Since 2003, hadron physics has entered a new era with the observation of a series of new hadronic states and the corresponding novel phenomena (see review articles~\cite{Chen:2016spr,Chen:2016qju,Liu:2019zoy} for more details), owing to the accumulation of experimental data with high precision. In 2003, the BaBar Collaboration observed a narrow state $D_{s0}^*(2317)$~\cite{Aubert:2003fg}, which decays into a $D_s^+\pi^0$ final state and has resonance parameters $m=2317.8\pm0.5$ MeV and $\Gamma<3.8$ MeV with spin parity $J^P=0^+$~\cite{Zyla:2020zbs}. Later, CLEO, Belle, and BaBar again confirmed this observation~\cite{Besson:2003cp,Aubert:2006bk,Aubert:2003pe,Aubert:2004pw,Abe:2003jk,Krokovny:2003zq}. Since its mass is about 100 MeV lower than the result of the quenched quark model~\cite{Godfrey:1985xj,Godfrey:2015dva}, there exists the so-called famous low-mass puzzle for $D_{s0}^*(2317)$. Such a situation not only results in the exotic state explanations including a hadronic $DK$ molecular state and compact tetraquark state proposed in Refs.~\cite{Barnes:2003dj,Faessler:2007gv,Xie:2010zza,Chen:2004dy,Dmitrasinovic:2005gc,Kim:2005gt}, but also stimulates theorists to pay more attention to the unquenched picture~\cite{vanBeveren:2003kd,Dai:2003yg,Hwang:2004cd,Simonov:2004ar,Ortega:2016mms,Cheng:2017oqh,Cheng:2014bca}, where the important role of the coupled-channel effect played in hadron spectroscopy starts to be realized. Later, low-mass puzzle phenomena appear in several other typical observed states $D_{s1}^\prime(2460)$~\cite{Cheng:2017oqh,Dai:2003yg}, $X(3872)$~\cite{Danilkin:2010cc,Li:2009ad,Kalashnikova:2005ui}, and $\Lambda_c(2940)$~\cite{Luo:2019qkm}, which naturally construct a complete chain from a heavy-light meson, and charmonium to heavy-light baryon, where the coupled-channel effect should be emphasized.

Under the unquenched picture for $D_{s0}^*(2317)$, $P$-wave bare state $D_s(0^+)$ can be dressed by the nearby $DK$ channel, which makes the physical mass be lowered down to be consistent with the mass of $D_{s0}^*(2317)$~\cite{vanBeveren:2003kd,Dai:2003yg,Hwang:2004cd,Simonov:2004ar,Ortega:2016mms,Cheng:2017oqh,Cheng:2014bca}. If replacing the antistrange quark $\bar s$ inside a  charmed-strange mesonic state by a $ss$ pair, we believe that there should exist a charmed-strange baryonic analog of $D_{s0}^*(2317)$, which inspires our interest in exploring whether the coupled-channel effect may play an important role in such a new system corresponding to the $P$-wave $\Omega_{c0}(1P,1/2^-)$ system.

\begin{figure}
\includegraphics[width=8.6cm,keepaspectratio]{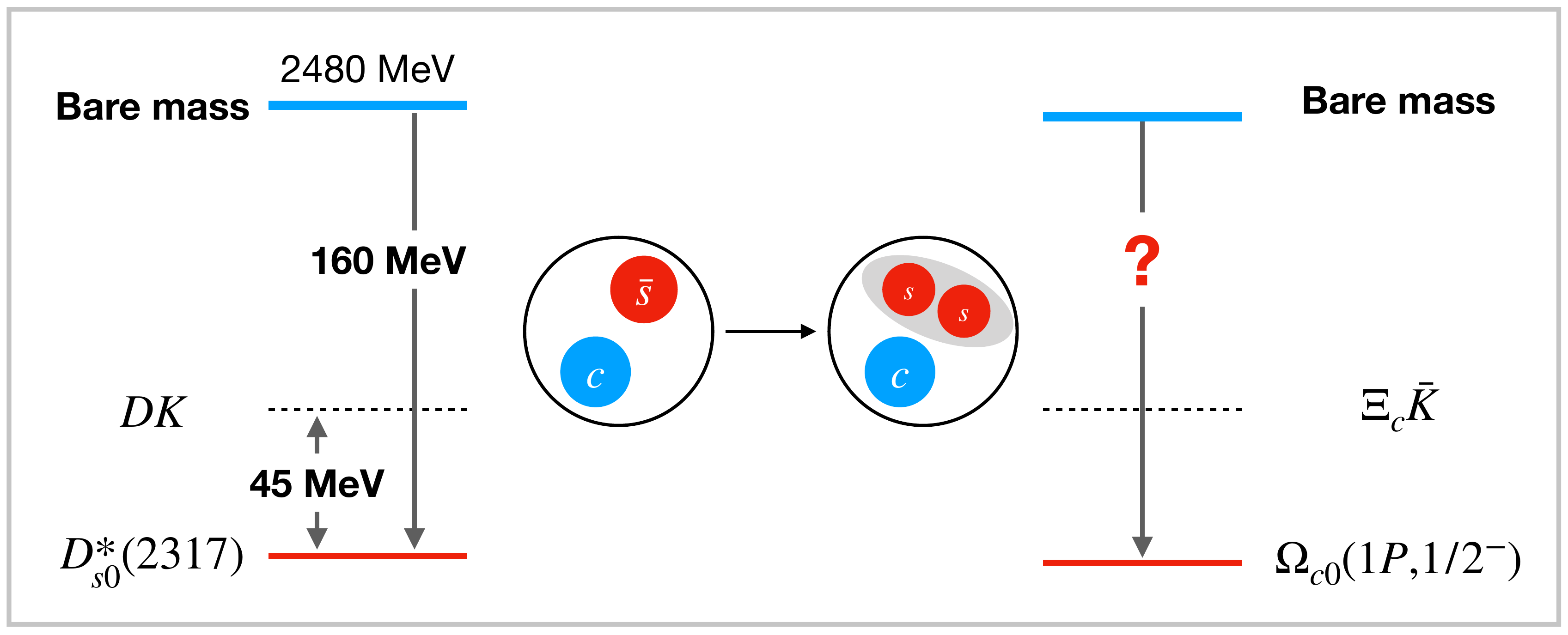}
\caption{The similarity between $D_{s0}^*(2317)$ and $\Omega_{c0}(1P,1/2^-)$. The predicted bare mass of $D_s$ meson is taken from the GI model~\cite{Godfrey:1985xj,Godfrey:2015dva}. The dashed lines represent $DK$ threshold (left) and $\Xi_c\bar{K}$ threshold (right).}
\label{analogycbarscss}
\end{figure}

As indicated in Fig.~\ref{analogycbarscss}, the bare mass of $\Omega_{c0}(1P,1/2^-)$ predicted by most of quenched models~\cite{Ebert:2007nw,Yoshida:2015tia,Shah:2016nxi,Maltman:1980er,Chen:2017gnu} is above the $\Xi_c\bar{K}$ threshold and there exists typical $S$-wave interaction between $\Omega_{c0}(1P,1/2^-)$ and the $\Xi_c\bar{K}$ channel. Thus, we have reason to believe that the coupled channel is obviously effective. In this work, we adopt an unquenched quark model to quantitatively reflect the existing coupled-channel effects. Before doing a realistic calculation for $\Omega_{c0}(1P,1/2^-)$, we first study $D_{s0}(2317)$ with the same framework, by which we can check the reliability of the adopted unquenched quark model. Our calculation explicitly shows that the coupled-channel effect on $D_s(0^+)$ can exactly reproduce the mass of $D_{s0}(2317)$. Naturally, when we continue to focus on $\Omega_{c0}(1P,1/2^-)$, we find that a low-mass phenomenon still exists, which makes the physical mass of $\Omega_{c0}(1P,1/2^-)$ lower than the $\Xi_c\bar{K}$ threshold. This fact further shows that $\Omega_{c0}(1P,1/2^-)$ should be a narrow state. The predicted behavior of $\Omega_{c0}(1P,1/2^-)$ as teh charmed-strange analog of $D_{s0}^*(2317)$ can be examined in future experiments.

In 2017, the LHCb and Belle collaborations have reported higher states for the $\Omega_c$ family~\cite{Aaij:2017nav,Yelton:2017qxg}. By taking this opportunity, we have a further discussion on the possible relation of the predicted charmed-strange baryonic analog of $D_{s0}^*(2317)$ and these observations.

This paper is organized as follows. After the introduction, we take the $D_{s0}^*(2317)$ as a sample to test the effectiveness of the unquenched model in Sec.~\ref{Ds2317}. Next, in Sec.~\ref{Omegac0}, we employ the same model to calculate the unquenched mass of $\Omega_{c0}(1P,1/2^-)$ with a coupled-channel effect from the $\Xi_c\bar{K}$ channel. Finally, the paper ends with the conclusions and discussions in Sec.~\ref{summary}.

\section{Test the Unquenched Model for $D_{s0}^*(2317)$}\label{Ds2317}

For a heavy-light hadron, the basis with total angular momentum $J$ in heavy quark symmetry is
\begin{equation}\label{sljellsQ}
|s_\ell,L,j_\ell,J\rangle=|[[s_\ell\otimes L]_{j_\ell}\otimes s_Q]_J\rangle,
\end{equation}
where $j_\ell$ and $s_Q$ are the angular momentum of light degree of freedom and spin of a heavy quark, respectively. The $s_\ell$ and $L$ in Eq.~(\ref{sljellsQ}) are the spin of light degree of freedom and orbital angular momentum, respectively. For the $D_s(1P)$ states, the light degree of freedom is $s_\ell=\frac{1}{2}$ and $L=1$, and, hence, the possible angular momenta of light degree of freedom are $j_\ell=\frac{1}{2}$ and $\frac{3}{2}$.

To make our conclusion for the $P$-wave $\Omega_{c0}(1P,1/2^-)$ state more reliable,  we test the adopted unquenched model in this section by examining the coupled-channel effect on $D_{s0}^*(2317)$. We not only illustrate why the mass of $D_{s0}^*(2317)$ shifts down about 80 MeV, but also fix the parameters of the unquenched quark model, which can be used to study the nontrivial coupled-channel effect on the  $P$-wave $\Omega_c$ states.

Because of the unquenched effect, the physical $D_{s0}^*(2317)$ state contains both $c\bar{s}$ and $DK$ components, which could be denoted as~\cite{Lu:2017hma,Anwar:2018yqm}
\begin{equation}\label{componentds2317}
|{D^*_{s0}(2317)}\rangle = c_{c\bar{s}}|c\bar{s}(1^3P_0)\rangle + \int {\rm d}^3 {\bf p}\;c_{DK}({\bf p})|{DK,{\bf p}}\rangle.
\end{equation}
Here, the $c_{c\bar{s}}$ denotes the probability amplitude of the $c\bar{s}$ core in the $D_{s0}^*(2317)$ wave function, and the $c_{DK}({\bf p})$ is the component of the $DK$ channel. The $c\bar{s}(1^3P_0)$ in Eq.~(\ref{componentds2317}) represents the conventional $D_s(0^+)$ with radial quantum number $n=0$ [see Eq.~(\ref{sho}) for the spatial wave function]. Then, the full Hamiltonian of the physical $D_{s0}^*(2317)$ state can be written as~\cite{Kalashnikova:2005ui,Liu:2011yp}
\begin{equation}\label{fullHamiltonian}
\hat{H} = \left(\begin{array}{cc}
          \hat{H}_0&\hat{H}_I\\
          \hat{H}_I&\hat{H}_{DK}
          \end{array}\right).
\end{equation}
The $\hat{H}_0$ is the Hamiltonian in a conventional quark model, by which one obtains the discrete mass spectrum of the bare charmed-strange mesons. The $\hat{H}_{DK}$ refers to the free Hamiltonian of the continuum states $|DK\rangle$, i.e.,
\begin{equation}\label{HDK}
\hat{H}_{DK}|{DK, {\bf p}}\rangle = \left(\sqrt{m_D^2+p^2}+\sqrt{m_K^2+p^2}\right)|{DK, {\bf p}}\rangle,
\end{equation}
where the interactions between the $D$ and $K$ mesons are neglected. The $\hat{H}_I$ that causes a mixture of the pure $c\bar{s}$ state (bare state) and $DK$ continuum can be borrowed from the quark-pair-creation (QPC or $^3P_0$) model~\cite{Micu:1968mk,LeYaouanc:1972vsx,LeYaouanc:1973ldf,Barnes:2007xu,Ackleh:1996yt}. In the nonrelativistic limit, the transition operator $\hat{H}_I$ can be expressed as
\begin{equation}\label{toperator}
\begin{split}
\hat{H}_I=&-3\gamma\sum_{m}\langle 1,m;1,-m|0,0\rangle\int{\rm d}^3{\bf p}_i\;{\rm d}^3{\bf p}_j\;\delta({\bf p}_i+{\bf p}_j)\\
          &\times\mathcal{Y}_1^m\left(\frac{{\bf p}_i-{\bf p}_j}{2}\right)\omega_0^{(i,j)}\phi_0^{(i,j)}\chi_{1,-m}^{(i,j)}b^\dagger_i({\bf p}_i)d^\dagger_j({\bf p}_j),
\end{split}
\end{equation}
where $\omega$, $\phi$, $\chi$ and $\mathcal{Y}$ are the color, flavor, spin, and spatial functions of the quark pair, respectively. The $b_{i}^\dagger$ and $d_{j}^\dagger$ are quark and antiquark creation operators, respectively. The dimensionless parameter $\gamma$ describes the strength of a quark-antiquark pair created from the vacuum. Now the amplitude of $c\bar{s}(1^3P_0)\to DK$ can be denoted as
\begin{equation}\label{amplitude}
{\cal M}_{c\bar{s}(1^3P_0)\to DK}(p)=\langle DK, {\bf p}|\hat{H}_I|c\bar{s}(1^3P_0)\rangle,
\end{equation}
where $p$ represents the momentum of $D$ meson in the center-of-mass frame of the $c\bar{s}(1^3P_0)$ state.

With the above preparation, the Schr\"{o}dinger equation for $D_{s0}^*(2317)$ could be denoted as
\begin{equation}\label{singleequ}
\left(\begin{array}{cc}
\hat{H}_0&\hat{H}_I\\
\hat{H}_I&\hat{H}_{DK}
\end{array}\right)
\left(\begin{array}{c}c_{c\bar{s}}|{c\bar{s}(1^3P_0)}\rangle\\c_{DK}|DK\rangle\end{array}\right)
=M
\left(\begin{array}{c}c_{c\bar{s}}|{c\bar{s}(1^3P_0)}\rangle\\c_{DK}|DK\rangle\end{array}\right).
\end{equation}
After diagonalization of Eq.~(\ref{singleequ}), we obtain the following coupled-channel equation
\begin{equation}\label{eqnmphy}
M-M_0-\Delta M(M)=0.
\end{equation}
$\Delta M(M)$ is the mass shift with definition
\begin{equation}\label{eqdeltam}
\Delta M(M)={\rm Re}\int_0^\infty p^2{\rm d}p\frac{|{\cal M}_{c\bar{s}(1^3P_0)\to DK}(p)|^2}{M-\sqrt{M_D^2+p^2}-\sqrt{M_K^2+p^2}}.
\end{equation}
The probability of $c\bar{s}$ could be determined by
\begin{equation}
|c_{c\bar{s}}|^2=\left(1-\left.\frac{\partial \Delta M(M)}{\partial M}\right|_{M=M^{\rm phy}}\right)^{-1},
\end{equation}
where the $M^{\rm phy}$ is the solution of Eqs.~(\ref{eqnmphy}) and (\ref{eqdeltam}). Since we consider only $c\bar s$ and $DK$ components, we could consider that $1-|c_{c\bar{s}}|^2$ is the probability of $DK$.

To extract the mass shift of $D_{s0}^*(2317)$ state by Eq.~(\ref{eqnmphy}), one should obtain the bare mass $M_0$ as the first step. In the following, we employ a nonrelativistic potential model to calculate the mass spectrum of the bare charmed-strange mesons. The Hamiltonian is given as
\begin{equation}\label{H0}
\hat{H}_0=\sum\limits_{i=1}\left(m_i+\frac{p_i^2}{2m_i}\right)+\sum\limits_{i<j}V_{ij},
\end{equation}
where $m_i$ and $p_i$ are the mass and momentum, respectively, of the $i$th constituent quark. The $V_{ij}$ in Eq.~(\ref{H0}) is the interaction between quark and quark (or quark and antiquark), which contains one-gluon-exchange (OGE) potentials and confining potentials and could be expanded as
\begin{equation}\label{Vij}
V_{ij}=H_{ij}^{\rm conf}+H_{ij}^{\rm hyp}+H_{ij}^{\rm so(cm)}+H_{ij}^{\rm so(tp)}.
\end{equation}
The first term of $V_{ij}$ in Eq.~(\ref{Vij}) is the Cornell potential, which is spin independent, i.e.,
\begin{equation}\label{conf1}
H_{ij}^{\rm conf}=-\frac{4}{3}\frac{\alpha_s}{r_{ij}}+br_{ij}+C,
\end{equation}
where the $\alpha_s$, $b$, and $C$ denote the coupling constant of OGE, the strength of linear confinement, and mass-renormalized constant, respectively. Besides the spin-independent term, $V_{ij}$ also contains spin-spin interaction, i.e., the hyperfine interaction is
\begin{equation}
H_{ij}^{\rm hyp}=\frac{4\alpha_s}{3m_im_j}\left(\frac{8\pi}{3}{\bf s}_i\cdot{\bf s}_j\tilde{\delta}(r)+\frac{1}{r_{ij}^3}S({\bf r}_{ij},{\bf s}_i,{\bf s}_j)\right),
\end{equation}
where
\begin{equation}
\tilde{\delta}(r)=\frac{\sigma^3}{\pi^{3/2}}{\rm e}^{-\sigma^2r^2}
\end{equation}
is a Gaussian smearing function with a smearing parameter $\sigma$, and
\begin{equation}
S({\bf r}_{ij},{\bf s}_i,{\bf s}_j)=\frac{3{\bf s}_i\cdot{\bf r}_{ij}{\bf s}_j\cdot{\bf r}_{ij}}{r_{ij}^2}-{\bf s}_i\cdot{\bf s}_j
\end{equation}
is a tensor operator. Besides, the color-magnetic term and Thomas-precession piece of the spin-orbit interactions could be expressed as
\begin{equation}
H_{ij}^{\rm so(cm)}=\frac{4\alpha_s}{3r_{ij}^3}\left(\frac{1}{m_i}+\frac{1}{m_j}\right)\left(\frac{{\bf s}_i}{m_i}+\frac{{\bf s}_j}{m_j}\right)\cdot {\bf L}
\end{equation}
and
\begin{equation}
H_{ij}^{\rm so(tp)}=-\frac{1}{2r_{ij}}\frac{\partial H_{ij}^{\rm conf}}{\partial r_{ij}}\left(\frac{{\bf s}_i}{m_i^2}+\frac{{\bf s}_j}{m_j^2}\right)\cdot {\bf L},
\end{equation}
respectively.

The parameters in the quenched quark potential model are fixed by the low-lying well-established $\pi$, $K$, $D$, and $D_s$ mesons, i.e., $m_{u/d}=0.370\;{\rm GeV}$, $m_{s}=0.600\;{\rm GeV}$, and $m_{c}=1.880\;{\rm GeV}$, $\alpha_s=0.578$, $b=0.144\;{\rm GeV}^2$, $\sigma =1.028\;{\rm GeV}$, and $C=-0.685\;{\rm GeV}$. Using the above parameters, the predicted masses of $1^1S_0$, $1^3S_1$, $1P_1$ ($j_\ell=3/2$), and $1^3P_2$ are well consistent with the measured masses of $D_s(1968)$, $D_s^*(2112)$, $D_{s1}(2536)$, and $D_{s2}^*(2573)$, respectively. However, the mass of $D_s(1^3P_0)$ is obtained to be 2441 MeV, which is about 76 MeV above the $DK$ threshold and 124 MeV larger than the measured mass of $D_{s0}^*(2317)$. Our result is similar to the previous works in Refs.~\cite{Godfrey:1985xj,Godfrey:2015dva,DiPierro:2001dwf,Matsuki:1997da,Matsuki:2007zza}.

\begin{table}[h!]
\centering
\caption{The $\beta$ values of mesons in units of GeV.}
\label{betavaluemeson}
\renewcommand\arraystretch{1.05}
\begin{tabular*}{86mm}{@{\extracolsep{\fill}}
cc@{\hskip\tabcolsep\vrule width 0.75pt\hskip\tabcolsep}
cc@{\hskip\tabcolsep\vrule width 0.75pt\hskip\tabcolsep}
cc}
\toprule[1.00pt]
\toprule[1.00pt]
States             &$\beta$ &States             &$\beta$ &States           &$\beta$ \\
\midrule[0.75pt]
$\pi$              &0.409   &$D(1S)$            &0.357   &$D_s(1S)$        &0.428   \\
$K$                &0.385   &$D^*(1S)$          &0.307   &$D_s^*(1S)$      &0.371   \\
&                           &$D(1P)$            &0.204   &$D_s(1P)$        &0.237   \\
\bottomrule[1pt]
\bottomrule[1pt]
\end{tabular*}
\end{table}

To incorporate the coupled-channel effect for the $D_{s0}^*(2317)$, we adopt a simple harmonic oscillator wave function to depict the spatial wave function of a meson, i.e.,
\begin{equation}\label{sho}
\begin{split}
\psi_{nlm}(\beta,{\bf P})=&\frac{(-1)^n(-{\mathrm i})^l}{\beta^{\frac{3}{2}+l}}\sqrt{\frac{2n!}{\Gamma(n+\ell+\frac{3}{2})}}L_{n}^{l+\frac{1}{2}}({P^2}/{\beta^2})\\
&\times {\mathrm e}^{-\frac{P^2}{2\beta^2}}P^l Y_{l m}(\Omega_P),
\end{split}
\end{equation}
where $n$, $l$, and $m$ are radial, orbital, and magnetic quantum numbers, respectively. Then, the spatial wave function overlap in Eq.~(\ref{amplitude}) can be calculated analytically. The parameter $\beta$ that denotes the distance scale in momentum space could be extracted from the potential model mentioned above. In Table~\ref{betavaluemeson}, we collect the obtained $\beta$ values. The remaining parameter $\gamma$ is the strength of a quark pair creation from the vacuum. For the $D_s$ meson, we determine $\gamma=4.1$ from the width of $D_{s2}^*(2573)$.

\begin{figure}
\includegraphics[width=8.6cm,keepaspectratio]{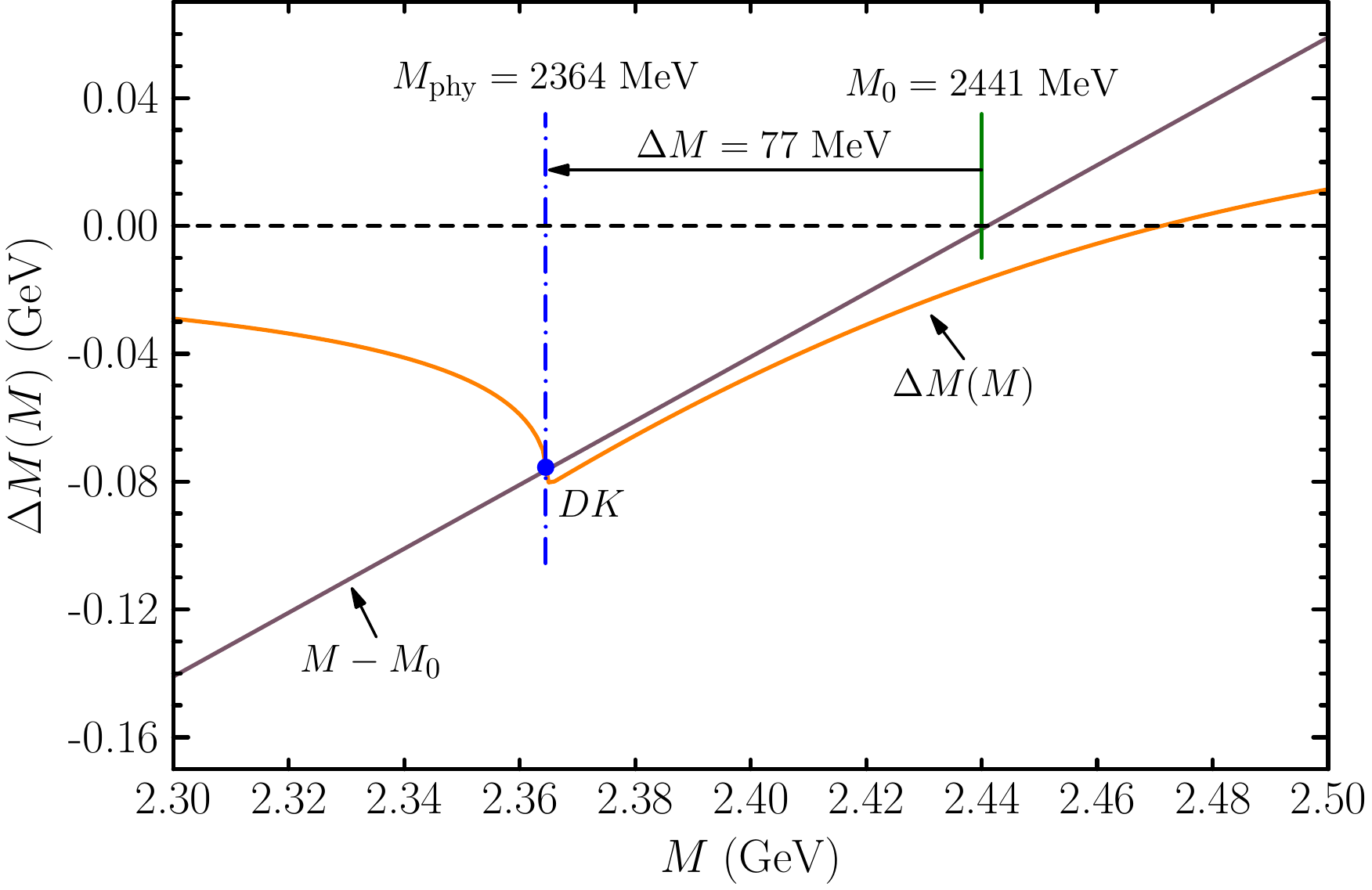}
\caption{The $M$ dependence of functions $M-M_0$ and $\Delta M(M)$ for $D_{s}(1^3P_0)$. Here, the $M_{\rm phy}$ value corresponding to the blue dot point at the intersection of two lines is the physical mass.}
\label{massDs2317}
\end{figure}

With the above preparations, we could calculate the physical mass of $D_s(1^3P_0)$. The numerical results are plotted in Fig.~\ref{massDs2317}. With the contribution from the intermediate $DK$ channel, the mass of the $D_s(1^3P_0)$ could be lowered down approximately to the experimental mass of $D_{s0}^*(2317)$. When the coupled-channel effect from $DK$ is considered, the mass of $D_s(1^3P_0)$ is shifted down from 2441 to 2364 MeV (a little below the $DK$ threshold), with a 77 MeV mass shift. The probabilities of $c\bar{s}$ and $DK$ are obtained to be about 16.6\% and 83.4\%, respectively. More importantly, the curve for $\Delta M(M)$ has a cusplike behavior at the $DK$ threshold, and the intersection between $M-M_0$ and $\Delta M(M)$ is very close to the cusplike position. The cusplike mass shift is a typical characteristic for nearby threshold states when including the coupled-channel effects from $S$-wave channels, which has been studied in many previous works~\cite{Li:2009ad,Tornqvist:1982yv,Isgur:1998kr,Luo:2019qkm,Tornqvist:1995kr,SilvestreBrac:1991pw}. Our results in Fig.~\ref{massDs2317} vividly describe how a nearby threshold state is affected by its $S$-wave channel.

\section{Prediction of a Charmed-strange Baryonic state $\Omega_{c0}(1P,1/2^-)$ as analog of $D_{s0}^*(2317)$}\label{Omegac0}

\begin{figure}[h!]
\includegraphics[width=8.6cm,keepaspectratio]{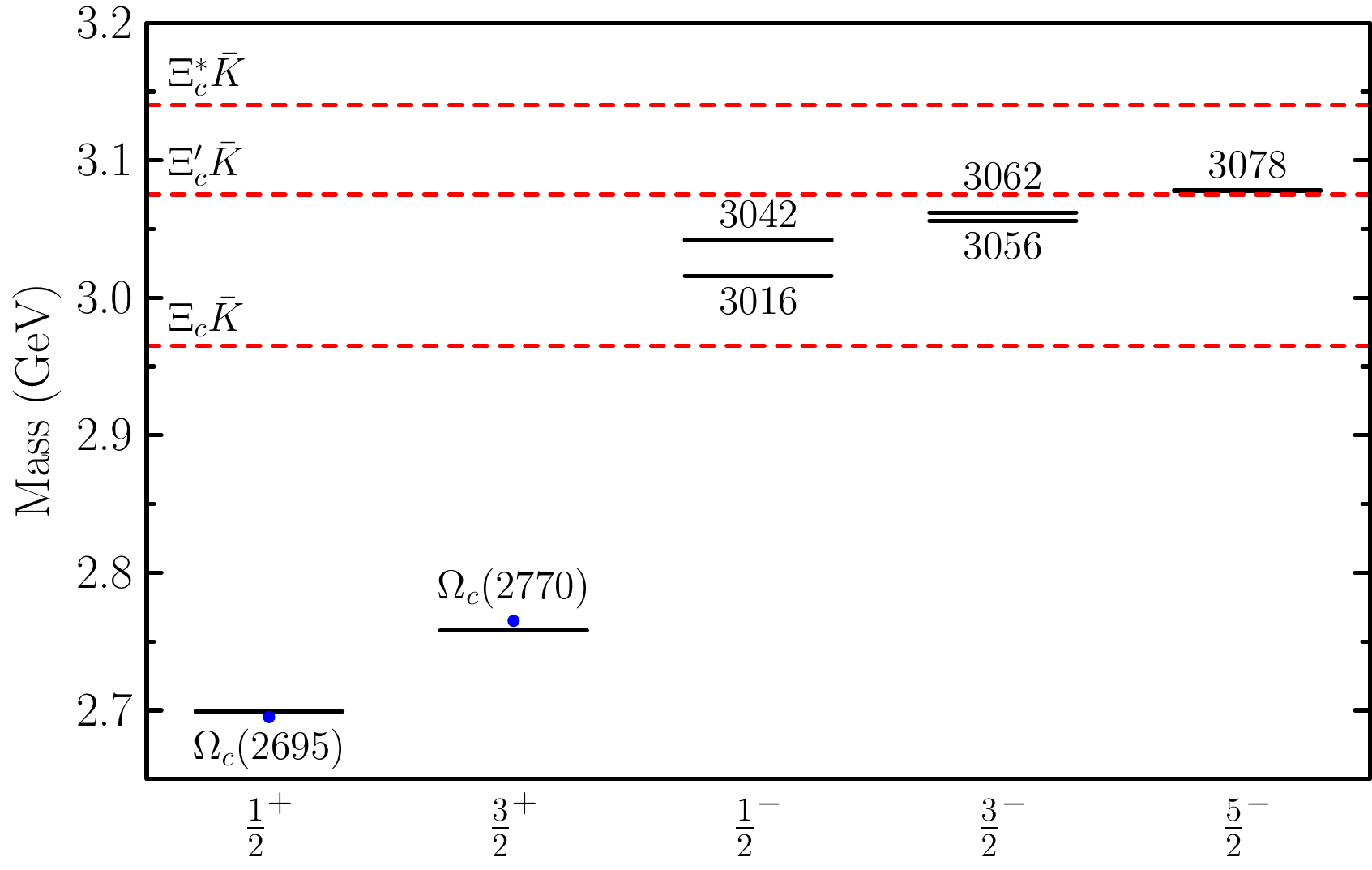}
\caption{The mass spectrum of $\Omega_c$ baryons. The short black lines denote the masses from the quenched quark model while the observed states are labeled by blue dots. The red dashed lines correspond to the thresholds.}
\label{massOmegacFig}
\end{figure}

In the following, we take the same unquenched quark model to predict the $\Omega_{c0}(1P,1/2^-)$ state below the threshold of $\Xi_c\bar{K}$. This state could be regarded as the charmed-strange baryonic analog of $D_{s0}^*(2317)$. To this end, one should first calculate the discrete mass spectrum of the bare $\Omega_c$ baryons. Here, we take the pairwise quark-quark potential to depict the spin-independent interactions of charm and strange quarks in the $\Omega_c$ system, i.e.,
\begin{equation}\label{VBij}
V=\sum_{i<j}\left(-\frac{2}{3}\frac{\alpha_s}{r_{ij}}+\frac{b}{2}r_{ij}+C\right).
\end{equation}
The interactions appearing in Eq.~(\ref{VBij}) can be regarded as the most direct way to extrapolate the interaction of meson [see Eq.~(\ref{conf1})] to the baryon system since the color factor $\langle{\bf F}_i\cdot {\bf F}_j\rangle$ of $qq$ in the baryon is $\frac{1}{2}$ that of $q\bar{q}$ in the meson. The color-magnetic term and Thomas-precession piece for the spin-orbit interactions of $P$-wave $\Omega_c$ baryons are taken from Ref.~\cite{Capstick:1986bm}, which are given by
\begin{equation}
\begin{split}
H_{ij}^{{\rm so(cm)}}=&\frac{2\alpha_s}{3r_{ij}^3}\left(\frac{{\bf r}_{ij}\times{\bf p}_i\cdot{\bf s}_i}{m_i^2}-\frac{{\bf r}_{ij}\times{\bf p}_j\cdot{\bf s}_j}{m_j^2}\right.\\
&\left.-\frac{{\bf r}_{ij}\times{\bf p}_j\cdot{\bf s}_i-{\bf r}_{ij}\times{\bf p}_i\cdot{\bf s}_j}{m_im_j}\right),\\
\end{split}
\end{equation}
and
\begin{equation}
\begin{split}
H_{ij}^{{\rm so(tp)}}=&-\frac{1}{2r_{ij}}\frac{\partial H_{ij}^{\rm conf}}{\partial r_{ij}}\left(\frac{{\bf r}_{ij}\times{\bf p}_i\cdot{\bf s}_i}{m_i^2}-\frac{{\bf r}_{ij}\times{\bf p}_j\cdot{\bf s}_j}{m_j^2}\right),
\end{split}
\end{equation}
respectively. The $\alpha_s$, $b$, and constituent quark masses for calculating the mass spectrum of $\Omega_c$ baryons have been fixed in the last section. Other parameters in the quark potential model are determined by the well-established $\Xi_c$ and $\Omega_c$ states, i.e., $\sigma=1.732\;{\rm GeV}$ and $C=-0.344\;{\rm GeV}$. Finally, we present the predicted masses of $\Omega_c$ baryons in Fig.~\ref{massOmegacFig}. The masses of $\Omega_c$ and $\Omega_c^*$ are fitted with the experimental results, and the mass of $\Omega_c(1P)$ is consistent with the previous predictions~\cite{Ebert:2007nw,Yoshida:2015tia,Shah:2016nxi,Maltman:1980er,Ebert:2011kk,Roberts:2007ni}.

In order to describe the wave function of a baryon conveniently, we employ the Jacobi coordinates ${\bm \rho}$ and ${\bm \lambda}$~\cite{Yoshida:2015tia} to depict the structure of a three-body system. In Table~\ref{betavaluebaryon}, we present the corresponding harmonic oscillator wave function scaling parameters $\beta_\rho$ and $\beta_\lambda$, which values are also fixed by the quark potential model. The remaining parameter is the $\gamma$ value in the QPC model. According to Ref.~\cite{Segovia:2012cd}, for different hadron systems, the $\gamma$ values are allowed to be different. In the charmed-strange baryon system, we determine $\gamma=8.66$ by the decay width of the well-established $\Xi_c(2790)$ state.

\begin{table}[htbp]
\centering
\caption{The $\beta$ values of baryons in units of GeV.}
\label{betavaluebaryon}
\renewcommand\arraystretch{1.05}
\begin{tabular*}{86mm}{@{\extracolsep{\fill}}
ccc@{\hskip\tabcolsep\vrule width 0.75pt\hskip\tabcolsep}
ccc@{\hskip\tabcolsep\vrule width 0.75pt\hskip\tabcolsep}
ccc}
\toprule[1.00pt]
\toprule[1.00pt]
States             &$\beta_\rho$ &$\beta_\lambda$ &States             &$\beta_\rho$ &$\beta_\lambda$ &States           &$\beta_\rho$ &$\beta_\lambda$ \\
\midrule[0.75pt]
$\Xi_c(1S)$        &0.313        &0.391           &$\Xi_c^\prime(1S)$ &0.256        &0.389           &$\Omega_c(1S)$   &0.289        &0.420           \\
$\cdots$           &$\cdots$     &$\cdots$        &$\Xi_c^*(1S)$      &0.245        &0.362           &$\Omega_c^*(1S)$ &0.276        &0.389           \\
$\Xi_c(1P)$        &0.292        &0.269           &$\Xi_c^\prime(1P)$ &0.243        &0.273           &$\Omega_c(1P)$   &0.273        &0.293           \\
\bottomrule[1pt]
\bottomrule[1pt]
\end{tabular*}
\end{table}

In the $\lambda$-mode-excited $\Omega_c(1P)$ states, $s_\ell=1$ is the spin of the $ss$ quark pair, and then the possible $j_\ell$ could be obtained as 0, 1, and 2. The $\Omega_c(1P)$ contain two $J^P=1/2^-$ states $|\frac{1}{2}^-\rangle_1$ and $|\frac{1}{2}^-\rangle_2$, which are the mixtures of $j_\ell=0$ and $j_\ell=1$ components, i.e.,
\begin{equation}\label{mixing1}
\left(\begin{array}{c}|\frac{1}{2}^-\rangle_1\\|\frac{1}{2}^-\rangle_2\end{array}\right)
=
\left(\begin{array}{cc}\cos\theta&-\sin\theta\\\sin\theta&\cos\theta\end{array}\right)
\left(\begin{array}{c}|j_\ell=0,\frac{1}{2}^-\rangle\\|j_\ell=1,\frac{1}{2}^-\rangle\end{array}\right),
\end{equation}
where $\theta$ is the mixing angle. One is a predominant $j_\ell=0$ state, while another is a predominant $j_\ell=1$ state. In the heavy quark limit, the pure $\Omega_c(1/2^-)_{j_\ell=0}$ state can couple with the $\Xi_c\bar{K}$, while the $\Omega_c(1/2^-)_{j_\ell=1}$ state is forbidden for the decay channel of $\Xi_c\bar{K}$. The mass of $J^P=1/2^-$ $\Omega_c(1P)$ with predominant $j_\ell=0$ is given by 3042 MeV in the quenched model, which is about 80 MeV higher than the $\Xi_c\bar{K}$ threshold. Since this $P$-wave $\Omega_c$ state mainly decays into $\Xi_c\bar{K}$ in the $S$-wave channel, its property is expected to be similar to the $D_{s0}^*(2317)$ state (see Fig.~\ref{analogycbarscss}). We also notice that the predicted mass of $\Omega_c(1/2^-)$ with predominant $j_\ell=1$ state is about 60 MeV below the threshold of its $S$-wave channel $\Xi^\prime_c\bar{K}$. Hence, the coupled-channel effect should also be considered for the predominant $j_\ell=1$ $\Omega_c(1/2^-)$ state.

Because of the special decay properties in the heavy quark limit, it is convenient to perform the calculation in the basis (so-called $j-j$ coupling scheme; see Ref.~\cite{Chen:2017gnu}). In the $j-j$ coupling scheme, the coupled-channel Schr\"{o}dinger equation which contains two bare states~\cite{Lu:2016mbb} could be written as
\begin{widetext}
\begin{equation}\label{multiequ}
\left(\begin{array}{cccc}
M^{j_\ell=0}&\tilde{V}^{\rm spin}&\int p^2{\rm d}{p}\langle\Omega_{c0}|\hat{H}_I|\Xi_c\bar{K}\rangle&0\\
\tilde{V}^{{\rm spin}}&M^{j_\ell=1}&0&\int p^2{\rm d}{p}\langle\Omega_{c1}|\hat{H}_I|\Xi_c^\prime\bar{K}\rangle\\
\langle \Xi_c\bar{K}|\hat{H}_I|\Omega_{c0}\rangle&0&H_{\Xi_c\bar{K}}&0\\
0&\langle \Xi_c^\prime\bar{K}|\hat{H}_I|\Omega_{c1}\rangle&0&H_{\Xi_c^{\prime}\bar{K}}
\end{array}\right)
\left(\begin{array}{c}c_0\\c_1\\c_{\Xi_c\bar{K}}\\c_{\Xi_c^{\prime}\bar{K}}\end{array}\right)
=M
\left(\begin{array}{c}c_0\\c_1\\c_{\Xi_c\bar{K}}\\c_{\Xi_c^{\prime}\bar{K}}\end{array}\right),
\end{equation}
\end{widetext}
where two $J^P=1/2^-$ $\Omega_c$ baryons with $j_\ell=0$ and $j_\ell=1$ are denoted as $\Omega_{c0}$ and $\Omega_{c1}$, respectively, and the effects of $S$-wave channels $\Xi_c\bar{K}$ and $\Xi^\prime_c\bar{K}$ are considered. The $M^{j_\ell=0}$ and $M^{j_\ell=1}$ are bare masses of $\Omega_{c0}$ and $\Omega_{c1}$, respectively, which can be obtained by the quark potential model. The off-diagonal element is defined as $\tilde{V}^{\rm spin}=\langle\Omega_{c0}|V^{\rm spin}|\Omega_{c1}\rangle$, which is hermitian and can be directly determined by the quark potential model. Finally, the multi-coupled-channel equation of Eq.~(\ref{multiequ}) can be simplified as (see details in the appendix~\ref{appendix})
\begin{equation}\label{multich}
\begin{split}
&\left(\begin{array}{cc}
M^{j_\ell=0}+\Delta M^0(M)&\tilde{V}^{\rm spin}\\
\tilde{V}^{{\rm spin}}&M^{j_\ell=1}+\Delta M^1(M)
\end{array}\right)
\left(\begin{array}{c}c_0\\c_1\end{array}\right)=
M\left(\begin{array}{c}c_0\\c_1\end{array}\right),
\end{split}
\end{equation}
where
\begin{equation}\label{DeltaM0}
\Delta M^0(M)={\rm Re}\int_0^\infty p^2{\rm d}p\frac{|{\cal M}_{\Omega_{c0}\to \Xi_c\bar{K}}(p)|^2}{M-\sqrt{M_{\Xi_c}^2+p^2}-\sqrt{M_K^2+p^2}},
\end{equation}
\begin{equation}\label{DeltaM1}
\Delta M^1(M)={\rm Re}\int_0^\infty p^2{\rm d}p\frac{|{\cal M}_{\Omega_{c1}\to \Xi_c^\prime\bar{K}}(p)|^2}{M-\sqrt{M_{\Xi_c^\prime}^2+p^2}-\sqrt{M_K^2+p^2}}.
\end{equation}
Equation (\ref{multich}) can be obviously decomposed into two independent single coupled-channel equations [as Eq.~(\ref{eqnmphy})] in the heavy quark limit $\tilde{V}^{\rm spin}\to 0$. Then, using Eqs.~(\ref{DeltaM0}) and (\ref{DeltaM1}), the probabilities of $\Xi_c\bar{K}$ and $\Xi_c^\prime\bar{K}$ could be written as
\begin{equation}\label{PXicKbar}
P_{\Xi_c\bar{K}}=-c_0^2\left.\frac{\partial \Delta M^0(M)}{\partial M}\right|_{M=M^{\rm phy}}
\end{equation}
and
\begin{equation}\label{PXicprimeKbar}
P_{\Xi_c^\prime\bar{K}}=-c_1^2\left.\frac{\partial \Delta M^1(M)}{\partial M}\right|_{M=M^{\rm phy}},
\end{equation}
respectively. In Eqs.~(\ref{PXicKbar}) and (\ref{PXicprimeKbar}), the values of $c_0$ and $c_1$ are extracted from the eigenvectors of Eq.~(\ref{multich}). Finally, with the condition
\begin{equation}
|c_0|^2+|c_1|^2+P_{\Xi_c\bar{K}}+P_{\Xi_c^\prime\bar{K}}=1,
\end{equation}
one could obtain the normalized values of $|c_0|^2$, $|c_1|^2$, $P_{\Xi_c\bar{K}}$, and $P_{\Xi_c^\prime\bar{K}}$.

By diagonalizing Eq.~({\ref{multich}}), the mass of $J^P=1/2^-$ $\Omega_c(1P)$ with predominant $j_\ell=0$ is predicted to be 2945 MeV, which is shifted down about 97 MeV. The mixing angle is simultaneously obtained as $\theta=\tan^{-1}\frac{c_1}{c_0}=-12.9^\circ$. Since the mixing angle is small, we tentatively call this state as $\Omega_{c0}^d(1P,1/2^-)$, where the superscript ``$d$" denotes that $\Omega_{c0}(1P,1/2^-)$ component is dominant. The probabilities of $\Xi_c\bar{K}$ and $\Xi_c^\prime\bar{K}$ are about 51.0\% and 0.3\%, respectively, and the probabilities of conventional $ssc$ components with $j_\ell=0$ and $j_\ell=1$ are about 46.2\% and 2.4\%, respectively. We would like to emphasize that the physical mass of $\Omega_{c0}^d(1P,1/2^-)$ becomes about 20 MeV below the $\Xi_c\bar{K}$ threshold when the nontrivial unquenched effect is incorporated. Then, we conclude that a charmed-strange baryonic analog of $D_{s0}^*(2317)$ may exist in the $P$-wave $\Omega_c$ baryon.

It is a problem how to search for the predicted $\Omega_{c0}^d(1P,1/2^-)$. If the mass of $\Omega_{c0}^d(1P,1/2^-)$ is below the $\Xi_c\bar{K}$ threshold, there is no OZI-allowed decay. The radiative decay channels $\Omega_{c0}^{(*)}\gamma$ and hadronic decay processes $\Omega_{c}^{(*)}\pi^0$ are kinematically allowed and should be considered in searches. 
{For the $\Omega_{c0}^d(1P,1/2^-)\to \Omega_{c}^{(*)}\pi^0$ decay, it is a typical
isospin-breaking process, where $\Omega_{c0}^d(1P,1/2^-)\to \Omega_{c}^{(*)}\pi^0$ may occur via $\eta-\pi^0$ mixing, which results in a suppression factor of $\sim10^{-4}$ \cite{Matsuki:2011xp}.}
Another approach of searching for $\Omega_{c0}(1P,1/2^-)$ is the radiative decay. Among the heavy flavor baryons, $\Xi_c^{\prime+}$, $\Xi_c^{\prime0}$, and $\Omega_c^*$ were discovered by radiative decays since their masses are below their respective lowest strong decay channels~\cite{Jessop:1998wt,Aubert:2006je,Solovieva:2008fw}. Very recently, the Belle Collaboration~\cite{Yelton:2020awh} has seen $\Xi_c(2790)^0$ and $\Xi_c(2815)^0$ in the radiative decay channel $\Xi_c^0\gamma$. This is a great breakthrough because the $\Xi_c^{\prime+}$, $\Xi_c^{\prime0}$, and $\Omega_c^*$ are $1S$ states, while $\Xi_c(2790)^0$ and $\Xi_c(2815)^0$ are orbitally excited states. In consideration of the fact that the excited states $\Xi_c(2790)$ and $\Xi_c(2815)$ can be seen via radiative decays, it is also probable to discover $\Omega_{c0}(1P,1/2^-)$ via $\Omega_c^{(*)}\gamma$ channels in future Belle II experiments.

The mass of another $J^P=1/2^-$ $\Omega_c$ state, i.e., the predominant $j_\ell=1$ state, is obtained as 2991 MeV by considering the coupled-channel effect from the $\Xi_c^\prime\bar{K}$ channel. Since this state is still above the $\Xi_c\bar{K}$, it is expected to be a conventional resonance. The mixing angle is determined as $\theta=-10.9^\circ$. Since Eq.~(\ref{multich}) is a multi-coupled-channel equation, it is not strange that the mixing angles for the two physical states contain a small difference~\cite{Lu:2016mbb,Fu:2018yxq}. Because the predicted mass is above the $\Xi_c\bar{K}$ threshold, the definitions of probabilities in Eqs.~(\ref{PXicKbar}) and (\ref{PXicprimeKbar}) may be not applicable in this situation~\cite{Baru:2003qq}. For a state above the threshold, the better description method is spectral density~\cite{Baru:2003qq,Kalashnikova:2005ui}. For completeness, we should also check the coupled-channel effect for two $J^P=3/2^-$ $\Omega_c(1P)$ states by the following relations
\begin{equation}\label{mixing2}
\left(\begin{array}{c}|\frac{3}{2}^-\rangle_1\\|\frac{3}{2}^-\rangle_2\end{array}\right)
=
\left(\begin{array}{cc}\cos\theta&-\sin\theta\\\sin\theta&\cos\theta\end{array}\right)
\left(\begin{array}{c}|j_\ell=1,\frac{3}{2}^-\rangle\\|j_\ell=2,\frac{3}{2}^-\rangle\end{array}\right).
\end{equation}
Our results indicate that the masses of two $J^P=3/2^-$ $\Omega_c$ states are not significantly affected by the coupled-channel effect. Their masses and mixing angles are given, respectively, by
\begin{equation}
\begin{split}
M_{\Omega_{c1}^d(1P,3/2^-)}^{\rm phy}=3029~{\rm MeV},~~~~~~~~&\theta=4.8^\circ;\\
M_{\Omega_{c2}^d(1P,3/2^-)}^{\rm phy}=3058~{\rm MeV},~~~~~~~~&\theta=4.1^\circ.
\end{split}
\end{equation}

\section{conclusions and discussions}\label{summary}

The observation of $D_{s0}^*(2317)$ makes theorists realize the importance of coupled-channel effects on mass spectrum study \cite{vanBeveren:2003kd,Dai:2003yg,Hwang:2004cd,Simonov:2004ar,Ortega:2016mms,Cheng:2017oqh,Cheng:2014bca}. If replacing the $\bar{s}$ quark in $D_{s0}^*(2317)$ by an $ss$ pair, we may naturally conjecture the existence of a new resonance as the charmed-strange baryonic analog of $D_{s0}^*(2317)$.
In this work, we have predicted such a new resonance by an unquenched quark model, where the predicted charmed-strange baryon has mass lower than the $\Xi_c\bar K$ threshold, and hence, its OZI-allowed strong decay mode is forbidden. Searching for this predicted charmed-strange baryon will be an interesting task for future experiments like Belle II and LHCb.

We have noticed that the LHCb Collaboration once reported five narrow $\Omega_c^0$ states, i.e., the $\Omega_c(3000)^0$, $\Omega_c(3050)^0$, $\Omega_c(3065)^0$, $\Omega_c(3090)^0$, and $\Omega_c(3120)^0$, in the $\Xi_c^+K^-$ channel~\cite{Aaij:2017nav}. The former four $\Omega_c^0$ states have been confirmed by the Belle Collaboration in the same decay channel, while the $\Omega_c(3120)^0$ signal has not been reported in Belle~\cite{Yelton:2017qxg}. These observed states are about in the range of 3.0$\sim$3.1 GeV, which is roughly fitted on the predicted mass region of conventional $\Omega_c(1P)$ states. Thus, some of the observed states could be good candidates of $\Omega_c(1P)$ states~\cite{Agaev:2017jyt,Karliner:2017kfm,Wang:2017zjw,Aliev:2017led,Cheng:2017ove,Zhao:2017fov,Wang:2017hej,
Chen:2017sci,Wang:2017vnc,Agaev:2017lip,Padmanath:2017lng,Chen:2017gnu}.
Besides the observed excited states above the $\Xi_c\bar{K}$ threshold, we think there should exist a missing $P$-wave state below the $\Xi_c\bar{K}$ threshold as suggested in this work. When the nontrivial coupled-channel effect has been considered, the mass of $J^P=1/2^-$ $\Omega_c(1P)$ with predominant $j_\ell=0$ should be shifted below the threshold of the $\Xi_c^+K^-$ channel. Obviously, this state cannot be found by the measured $\Xi_c^+K^-$ invariant mass spectrum from LHCb and Belle \cite{Aaij:2017nav,Yelton:2017qxg}. How to find this predicted charmed-strange baryon will be a challenging opportunity for the Belle II experiment, where this predicted charmed-strange baryon $\Omega_{c0}^d(1P,1/2^-)$ should decay into $\Omega_c^{(*)}\gamma$.

Before the present study, there exist several typical examples including $D_{s0}^*(2317)$, $D_{s0}^\prime(2460)$, $X(3872)$, and $\Lambda_c(2940)$, where the coupled channel may play a crucial role to understand their low-mass phenomena \cite{vanBeveren:2003kd,Dai:2003yg,Hwang:2004cd,Simonov:2004ar,Ortega:2016mms,Cheng:2017oqh,Cheng:2014bca,Danilkin:2010cc,Li:2009ad,Kalashnikova:2005ui,Luo:2019qkm}. If the predicted charmed-strange baryon as the charmed-strange baryonic analog of $D_{s0}^*(2317)$ can be confirmed in future experiments, it can provide a new example to show the importance of the coupled-channel effect.

\section*{Acknowledgments}

This work is supported by the China National Funds for Distinguished Young Scientists under Grant No. 11825503, National Key Research and Development Program of China under Contract No. 2020YFA0406400, the 111 Project under Grant No. B20063, and the National Natural Science Foundation of China under Grant No. 12047501. B. C. is partly supported by the National Natural Science Foundation of China under Grants No. 11305003 and No. 11647301.

\appendix
\section{}\label{appendix}
In this Appendix, we present some details of how to obtain Eq.~(\ref{multich}) from Eq.~(\ref{multiequ}). By expanding Eq.~(\ref{multiequ}), we obtain
\begin{equation}\label{omegac01}
\begin{split}
c_0 M^{j_\ell=0}+c_1\tilde{V}^{\rm spin}+\int p^2{\rm d}pc_{\Xi_c\bar{K}}\langle\Omega_{c0}|\hat{H}_I|\Xi_c\bar{K}\rangle=&c_0 M,\\
c_0\tilde{V}^{\rm spin}+c_1 M^{j_\ell=1}+\int p^2{\rm d}pc_{\Xi_c^\prime\bar{K}}\langle\Omega_{c1}|\hat{H}_I|\Xi_c^\prime\bar{K}\rangle=&c_1M
\end{split}
\end{equation}
and
\begin{equation}\label{cxicpk}
\begin{split}
c_0\langle \Xi_c\bar{K}|\hat{H}_I|\Omega_{c0}\rangle+c_{\Xi_c\bar{K}}H_{\Xi_c\bar{K}}=&c_{\Xi_c\bar{K}}M,\\
c_1\langle \Xi_c^\prime\bar{K}|\hat{H}_I|\Omega_{c1}\rangle+c_{\Xi_c^\prime\bar{K}}H_{\Xi_c^{\prime}\bar{K}}=&c_{\Xi_c^\prime\bar{K}}M.\\
\end{split}
\end{equation}
Using Eq.~(\ref{cxicpk}), we have
\begin{equation}
\begin{split}
c_{\Xi_c\bar{K}}=c_0\frac{\langle \Xi_c\bar{K}|\hat{H}_I|\Omega_{c0}\rangle}{M-H_{\Xi_c\bar{K}}},\;
c_{\Xi_c^\prime\bar{K}}=c_1\frac{\langle \Xi_c^\prime\bar{K}|\hat{H}_I|\Omega_{c1}\rangle}{M-H_{\Xi_c^{\prime}\bar{K}}}.
\end{split}
\end{equation}
Then Eq.~(\ref{omegac01}) could be rewritten as
\begin{equation}
\begin{split}
c_0 M^{j_\ell=0}+c_0\int p^2{\rm d}p\frac{|\langle \Xi_c\bar{K}|\hat{H}_I|\Omega_{c0}\rangle|^2}{M-H_{\Xi_c\bar{K}}}+c_1\tilde{V}^{\rm spin}=&c_0 M,\\
c_0\tilde{V}^{\rm spin}+c_1 M^{j_\ell=1}+c_1\int p^2{\rm d}p\frac{|\langle \Xi_c^\prime\bar{K}|\hat{H}_I|\Omega_{c1}\rangle|^2}{M-H_{\Xi_c^{\prime}\bar{K}}}=&c_1 M.
\end{split}
\end{equation}
The above relations are equivalent to the following eigenvalue equation
\begin{equation}
\begin{split}
&\left(\begin{array}{cc}
M^{j_\ell=0}+\Delta M^0(M)&\tilde{V}^{\rm spin}\\
\tilde{V}^{{\rm spin}}&M^{j_\ell=1}+\Delta M^1(M)
\end{array}\right)
\left(\begin{array}{c}c_0\\c_1\end{array}\right)=
M\left(\begin{array}{c}c_0\\c_1\end{array}\right),
\end{split}
\end{equation}
where
\begin{equation}
\Delta M^0(M)={\rm Re}\int_0^\infty p^2{\rm d}p\frac{|{\cal M}_{\Omega_{c0}\to \Xi_c\bar{K}}(p)|^2}{M-\sqrt{M_{\Xi_c}^2+p^2}-\sqrt{M_K^2+p^2}},
\end{equation}
\begin{equation}
\Delta M^1(M)={\rm Re}\int_0^\infty p^2{\rm d}p\frac{|{\cal M}_{\Omega_{c1}\to \Xi_c^\prime\bar{K}}(p)|^2}{M-\sqrt{M_{\Xi_c^\prime}^2+p^2}-\sqrt{M_K^2+p^2}}.
\end{equation}


\begin{thebibliography}{}
\bibitem{Chen:2016spr}
  H.~X.~Chen, W.~Chen, X.~Liu, Y.~R.~Liu and S.~L.~Zhu,
  A review of the open charm and open bottom systems,
  Rept.\ Prog.\ Phys.\  {\bf 80}, 076201 (2017).

\bibitem{Chen:2016qju}
  H.~X.~Chen, W.~Chen, X.~Liu and S.~L.~Zhu,
  The hidden-charm pentaquark and tetraquark states,
  Phys.\ Rept.\  {\bf 639} (2016) 1.

\bibitem{Liu:2019zoy}
  Y.~R.~Liu, H.~X.~Chen, W.~Chen, X.~Liu and S.~L.~Zhu,
  Pentaquark and Tetraquark states,
  Prog.\ Part.\ Nucl.\ Phys.\  {\bf 107}, 237 (2019).

\bibitem{Aubert:2003fg}
  B.~Aubert {\it et al.} [BaBar Collaboration],
  Observation of a narrow meson decaying to $D_s^+ \pi^0$ at a mass of 2.32 GeV$/c^2$,
  Phys.\ Rev.\ Lett.\  {\bf 90}, 242001 (2003).

\bibitem{Zyla:2020zbs}
  P.~A.~Zyla {\it et al.} [Particle Data Group],
  Review of Particle Physics,
  PTEP {\bf 2020}, no. 8, 083C01 (2020).

\bibitem{Besson:2003cp}
  D.~Besson {\it et al.} [CLEO Collaboration],
  Observation of a narrow resonance of mass 2.46 GeV$/c^2$ decaying to $D_s^{*+}\pi^0$ and confirmation of the $D^*_{sJ}(2317)$ state,
  Phys.\ Rev.\ D {\bf 68}, 032002 (2003)
  Erratum: [Phys.\ Rev.\ D {\bf 75}, 119908 (2007)].

\bibitem{Aubert:2006bk}
  B.~Aubert {\it et al.} [BaBar Collaboration],
  A Study of the $D^*_{sJ}(2317)$ and $D_{sJ}(2460)$ Mesons in Inclusive $c\bar{c}$ Production Near $\sqrt{s} = 10.6$ GeV,
  Phys.\ Rev.\ D {\bf 74}, 032007 (2006).

\bibitem{Aubert:2003pe}
  B.~Aubert {\it et al.} [BaBar Collaboration],
  Observation of a narrow meson decaying to $D_s^+ \pi^0 \gamma$ at a mass of 2.458 GeV/$c^2$,
  Phys.\ Rev.\ D {\bf 69}, 031101 (2004).

\bibitem{Aubert:2004pw}
  B.~Aubert {\it et al.} [BaBar Collaboration],
  Study of $B \to D_{sJ}^{(*)+} \bar{D}^{(*)}$ decays,
  Phys.\ Rev.\ Lett.\  {\bf 93}, 181801 (2004).

\bibitem{Abe:2003jk}
  Y.~Mikami {\it et al.} [Belle Collaboration],
  Measurements of the $D_{sJ}$ resonance properties,
  Phys.\ Rev.\ Lett.\  {\bf 92}, 012002 (2004).

\bibitem{Krokovny:2003zq}
  P.~Krokovny {\it et al.} [Belle Collaboration],
  Observation of the $D_{sJ}(2317)$ and $D_{sJ}(2457)$ in $B$ decays,
  Phys.\ Rev.\ Lett.\  {\bf 91}, 262002 (2003).

\bibitem{Godfrey:1985xj}
  S.~Godfrey and N.~Isgur,
  Mesons in a Relativized Quark Model with Chromodynamics,
  Phys.\ Rev.\ D {\bf 32}, 189 (1985).

\bibitem{Godfrey:2015dva}
  S.~Godfrey and K.~Moats,
  Properties of Excited Charm and Charm-Strange Mesons,
  Phys.\ Rev.\ D {\bf 93}, no. 3, 034035 (2016).

\bibitem{Barnes:2003dj}
  T.~Barnes, F.~E.~Close and H.~J.~Lipkin,
  Implications of a $DK$ molecule at 2.32 GeV,
  Phys.\ Rev.\ D {\bf 68}, 054006 (2003).

\bibitem{Faessler:2007gv}
  A.~Faessler, T.~Gutsche, V.~E.~Lyubovitskij and Y.~L.~Ma,
  Strong and radiative decays of the $D_{s0}^*(2317)$ meson in the $DK$-molecule picture,
  Phys.\ Rev.\ D {\bf 76}, 014005 (2007).

\bibitem{Xie:2010zza}
  Z.~X.~Xie, G.~Q.~Feng and X.~H.~Guo,
  Analyzing $D_{s0}^*(2317)^+$ in the $DK$ molecule picture in the Beth-Salpeter approach,
  Phys.\ Rev.\ D {\bf 81}, 036014 (2010).

\bibitem{Chen:2004dy}
  Y.~Q.~Chen and X.~Q.~Li,
  A Comprehensive four-quark interpretation of $D_s(2317)$, $D_s(2457)$ and $D_s(2632)$,
  Phys.\ Rev.\ Lett.\  {\bf 93}, 232001 (2004).

\bibitem{Dmitrasinovic:2005gc}
  V.~Dmitrasinovic,
  $D_{s0}^+(2317)$-$D_0(2308)$ mass difference as evidence for tetraquarks,
  Phys.\ Rev.\ Lett.\  {\bf 94}, 162002 (2005).

\bibitem{Kim:2005gt}
  H.~Kim and Y.~Oh,
  $D_s(2317)$ as a four-quark state in QCD sum rules,
  Phys.\ Rev.\ D {\bf 72}, 074012 (2005).

\bibitem{vanBeveren:2003kd}
  E.~van Beveren and G.~Rupp,
  Observed $D_s(2317)$ and tentative $D(2030)$ as the charmed cousins of the light scalar nonet,
  Phys.\ Rev.\ Lett.\  {\bf 91}, 012003 (2003).

\bibitem{Hwang:2004cd}
  D.~S.~Hwang and D.~W.~Kim,
  Mass of $D_{sJ}^*(2317)$ and coupled channel effect,
  Phys.\ Lett.\ B {\bf 601}, 137 (2004).

\bibitem{Simonov:2004ar}
  Y.~A.~Simonov and J.~A.~Tjon,
  The Coupled-channel analysis of the $D$ and $D_s$ mesons,
  Phys.\ Rev.\ D {\bf 70}, 114013 (2004).

\bibitem{Ortega:2016mms}
  P.~G.~Ortega, J.~Segovia, D.~R.~Entem and F.~Fernandez,
  Molecular components in $P$-wave charmed-strange mesons,
  Phys.\ Rev.\ D {\bf 94}, no. 7, 074037 (2016).

\bibitem{Cheng:2017oqh}
  H.~Y.~Cheng and F.~S.~Yu,
  Masses of Scalar and Axial-Vector B Mesons Revisited,
  Eur.\ Phys.\ J.\ C {\bf 77}, no. 10, 668 (2017).

\bibitem{Cheng:2014bca}
  H.~Y.~Cheng and F.~S.~Yu,
  Near mass degeneracy in the scalar meson sector: Implications for $B^*_{(s)0}$ and $B'_{(s)1}$ mesons,
  Phys.\ Rev.\ D {\bf 89}, no. 11, 114017 (2014).

\bibitem{Dai:2003yg}
  Y.~B.~Dai, C.~S.~Huang, C.~Liu and S.~L.~Zhu,
  Understanding the $D^+_{sJ}(2317)$ and $D^+_{sJ}(2460)$ with sum rules in HQET,
  Phys.\ Rev.\ D {\bf 68}, 114011 (2003).

\bibitem{Li:2009ad}
  B.~Q.~Li, C.~Meng and K.~T.~Chao,
  Coupled-Channel and Screening Effects in Charmonium Spectrum,
  Phys.\ Rev.\ D {\bf 80}, 014012 (2009).

\bibitem{Kalashnikova:2005ui}
  Y.~S.~Kalashnikova,
  Coupled-channel model for charmonium levels and an option for $X(3872)$,
  Phys.\ Rev.\ D {\bf 72}, 034010 (2005).


\bibitem{Danilkin:2010cc}
  I.~V.~Danilkin and Y.~A.~Simonov,
  Dynamical origin and the pole structure of $X(3872)$,
  Phys.\ Rev.\ Lett.\  {\bf 105}, 102002 (2010).

\bibitem{Luo:2019qkm}
  S.~Q.~Luo, B.~Chen, Z.~W.~Liu and X.~Liu,
  Resolving the low mass puzzle of $\Lambda_c(2940)^+$,
  Eur.\ Phys.\ J.\ C {\bf 80}, no. 4, 301 (2020).

\bibitem{Ebert:2007nw}
  D.~Ebert, R.~N.~Faustov and V.~O.~Galkin,
  Masses of excited heavy baryons in the relativistic quark model,
  Phys.\ Lett.\ B {\bf 659}, 612 (2008).

\bibitem{Yoshida:2015tia}
  T.~Yoshida, E.~Hiyama, A.~Hosaka, M.~Oka and K.~Sadato,
  Spectrum of heavy baryons in the quark model,
  Phys.\ Rev.\ D {\bf 92}, no. 11, 114029 (2015).

\bibitem{Shah:2016nxi}
  Z.~Shah, K.~Thakkar, A.~K.~Rai and P.~C.~Vinodkumar,
  Mass spectra and Regge trajectories of $\Lambda_{c}^{+}$, $\Sigma_{c}^{0}$, $\Xi_{c}^{0}$ and $\Omega_{c}^{0}$ baryons,
  Chin.\ Phys.\ C {\bf 40}, no. 12, 123102 (2016).

\bibitem{Chen:2017gnu}
  B.~Chen and X.~Liu,
  New $\Omega_c^0$ baryons discovered by LHCb as the members of $1P$ and $2S$ states,
  Phys.\ Rev.\ D {\bf 96}, no. 9, 094015 (2017).

\bibitem{Maltman:1980er}
  K.~Maltman and N.~Isgur,
  Baryons With Strangeness and Charm in a Quark Model With Chromodynamics,
  Phys.\ Rev.\ D {\bf 22}, 1701 (1980).

\bibitem{Aaij:2017nav}
  R.~Aaij {\it et al.} [LHCb Collaboration],
  Observation of five new narrow $\Omega_c^0$ states decaying to $\Xi_c^+ K^-$,
  Phys.\ Rev.\ Lett.\  {\bf 118}, no. 18, 182001 (2017).

\bibitem{Yelton:2017qxg}
  J.~Yelton {\it et al.} [Belle Collaboration],
  Observation of Excited $\Omega_c$ Charmed Baryons in $e^+e^-$ Collisions,
  Phys.\ Rev.\ D {\bf 97}, no. 5, 051102 (2018).

\bibitem{Lu:2017hma}
  Y.~Lu, M.~N.~Anwar and B.~S.~Zou,
  How Large is the Contribution of Excited Mesons in Coupled-Channel Effects?
  Phys.\ Rev.\ D {\bf 95}, no. 3, 034018 (2017).

\bibitem{Anwar:2018yqm}
  M.~N.~Anwar, Y.~Lu and B.~S.~Zou,
  $\chi_{b}(3P)$ multiplet revisited: Hyperfine mass splitting and radiative transitions,
  Phys.\ Rev.\ D {\bf 99}, no. 9, 094005 (2019).

\bibitem{Liu:2011yp}
  J.~F.~Liu and G.~J.~Ding,
  Bottomonium Spectrum with Coupled-Channel Effects,
  Eur.\ Phys.\ J.\ C {\bf 72}, 1981 (2012).

\bibitem{Micu:1968mk}
  L.~Micu,
  Decay rates of meson resonances in a quark model,
  Nucl.\ Phys.\ B {\bf 10}, 521 (1969).

\bibitem{LeYaouanc:1972vsx}
  A.~Le Yaouanc, L.~Oliver, O.~Pene and J.~C.~Raynal,
  Naive quark pair creation model of strong interaction vertices,
  Phys.\ Rev.\ D {\bf 8}, 2223 (1973).

\bibitem{LeYaouanc:1973ldf}
  A.~Le Yaouanc, L.~Oliver, O.~Pene and J.-C.~Raynal,
  Naive quark pair creation model and baryon decays,
  Phys.\ Rev.\ D {\bf 9}, 1415 (1974).

\bibitem{Barnes:2007xu}
  T.~Barnes and E.~S.~Swanson,
  Hadron loops: General theorems and application to charmonium,
  Phys.\ Rev.\ C {\bf 77}, 055206 (2008).

\bibitem{Ackleh:1996yt}
  E.~S.~Ackleh, T.~Barnes and E.~S.~Swanson,
  On the mechanism of open flavor strong decays,
  Phys.\ Rev.\ D {\bf 54}, 6811 (1996).

\bibitem{DiPierro:2001dwf}
  M.~Di Pierro and E.~Eichten,
  Excited Heavy - Light Systems and Hadronic Transitions,
  Phys.\ Rev.\ D {\bf 64}, 114004 (2001).

\bibitem{Matsuki:1997da}
  T.~Matsuki and T.~Morii,
  Spectroscopy of heavy mesons expanded in $1/m_Q$,
  Phys.\ Rev.\ D {\bf 56}, 5646 (1997).

\bibitem{Matsuki:2007zza}
  T.~Matsuki, T.~Morii and K.~Sudoh,
  New heavy-light mesons $Q\bar{q}$,
  Prog.\ Theor.\ Phys.\  {\bf 117}, 1077 (2007).

\bibitem{Tornqvist:1995kr}
  N.~A.~Tornqvist,
  Understanding the scalar meson $q\bar{q}$ nonet,
  Z.\ Phys.\ C {\bf 68}, 647 (1995).

\bibitem{Tornqvist:1982yv}
  N.~A.~Tornqvist,
  Scalar Mesons in the Unitarized Quark Model,
  Phys.\ Rev.\ Lett.\  {\bf 49}, 624 (1982).

\bibitem{Isgur:1998kr}
  N.~Isgur,
  Spin orbit inversion of excited heavy quark mesons,
  Phys.\ Rev.\ D {\bf 57}, 4041 (1998).



\bibitem{SilvestreBrac:1991pw}
  B.~Silvestre- Brac and C.~Gignoux,
  Unitary effects in spin orbit splitting of $P$-wave baryons,
  Phys.\ Rev.\ D {\bf 43}, 3699 (1991).



\bibitem{Capstick:1986bm}
  S.~Capstick and N.~Isgur,
  Baryons in a Relativized Quark Model with Chromodynamics,
  Phys.\ Rev.\ D {\bf 34}, 2809 (1986)
  [AIP Conf.\ Proc.\  {\bf 132}, 267 (1985)].

\bibitem{Ebert:2011kk}
  D.~Ebert, R.~N.~Faustov and V.~O.~Galkin,
  Spectroscopy and Regge trajectories of heavy baryons in the relativistic quark-diquark picture,
  Phys.\ Rev.\ D {\bf 84}, 014025 (2011).

\bibitem{Roberts:2007ni}
  W.~Roberts and M.~Pervin,
  Heavy baryons in a quark model,
  Int.\ J.\ Mod.\ Phys.\ A {\bf 23}, 2817 (2008).

\bibitem{Segovia:2012cd}
  J.~Segovia, D.~R.~Entem and F.~Fernandez,
  Scaling of the $^3P_0$ Strength in Heavy Meson Strong Decays,
  Phys.\ Lett.\ B {\bf 715}, 322 (2012).

\bibitem{Matsuki:2011xp}
  T.~Matsuki and K.~Seo,
  Chiral Particle Decay of Heavy-Light Mesons in a Relativistic Potential Model,
  Phys.\ Rev.\ D {\bf 85}, 014036 (2012).

\bibitem{Jessop:1998wt}
  C.~P.~Jessop {\it et al.} [CLEO Collaboration],
  Observation of two narrow states decaying into $\Xi^+_c \gamma$ and $\Xi^0_c \gamma$,
  Phys.\ Rev.\ Lett.\  {\bf 82}, 492 (1999).

\bibitem{Aubert:2006je}
  B.~Aubert {\it et al.} [BaBar Collaboration],
  Observation of an excited charm baryon $\Omega_c^*$ decaying to $\Omega_c^0 \gamma$,
  Phys.\ Rev.\ Lett.\  {\bf 97}, 232001 (2006).

\bibitem{Solovieva:2008fw}
  E.~Solovieva {\it et al.},
  Study of $\Omega_c^{0}$ and $\Omega_c^{*0}$ Baryons at Belle,
  Phys.\ Lett.\ B {\bf 672}, 1 (2009).

\bibitem{Yelton:2020awh}
  J.~Yelton {\it et al.} [Belle Collaboration],
  Study of electromagnetic decays of orbitally excited $\Xi_c$ baryons,
  Phys.\ Rev.\ D {\bf 102}, no. 7, 071103 (2020).

\bibitem{Lu:2016mbb}
  Y.~Lu, M.~N.~Anwar and B.~S.~Zou,
  Coupled-Channel Effects for the Bottomonium with Realistic Wave Functions,
  Phys.\ Rev.\ D {\bf 94}, no. 3, 034021 (2016).

\bibitem{Fu:2018yxq}
  H.~F.~Fu and L.~Jiang,
  Coupled-channel-induced $S{-}D$ mixing of Charmonia and testing possible assignments for $Y$(4260) and $Y$(4360),
  Eur.\ Phys.\ J.\ C {\bf 79}, no. 6, 460 (2019).

\bibitem{Baru:2003qq}
  V.~Baru, J.~Haidenbauer, C.~Hanhart, Y.~Kalashnikova and A.~E.~Kudryavtsev,
  Evidence that the $a_0(980)$ and $f_0(980)$ are not elementary particles,
  Phys.\ Lett.\ B {\bf 586}, 53 (2004).

\bibitem{Agaev:2017jyt}
  S.~S.~Agaev, K.~Azizi and H.~Sundu,
  On the nature of the newly discovered $\Omega_c$ states,
  EPL {\bf 118}, no. 6, 61001 (2017).

\bibitem{Karliner:2017kfm}
  M.~Karliner and J.~L.~Rosner,
  Very narrow excited $\Omega_c$ baryons,
  Phys.\ Rev.\ D {\bf 95}, no. 11, 114012 (2017).

\bibitem{Wang:2017zjw}
  Z.~G.~Wang,
  Analysis of $\Omega _c(3000)$ , $\Omega _c(3050)$ , $\Omega _c(3066)$ , $\Omega _c(3090)$ and $\Omega _c(3119)$ with QCD sum rules,
  Eur.\ Phys.\ J.\ C {\bf 77}, no. 5, 325 (2017).

\bibitem{Cheng:2017ove}
  H.~Y.~Cheng and C.~W.~Chiang,
  Quantum numbers of $\Omega_c$ states and other charmed baryons,
  Phys.\ Rev.\ D {\bf 95}, no. 9, 094018 (2017).

\bibitem{Zhao:2017fov}
  Z.~Zhao, D.~D.~Ye and A.~Zhang,
  Hadronic decay properties of newly observed $\Omega_c$ baryons,
  Phys.\ Rev.\ D {\bf 95}, no. 11, 114024 (2017).

\bibitem{Aliev:2017led}
  T.~M.~Aliev, S.~Bilmis and M.~Savci,
  Are the new excited $\Omega_c$ baryons negative parity states?
  Mod.\ Phys.\ Lett.\ A {\bf 35}, no. 01, 1950344 (2020).

\bibitem{Wang:2017hej}
  K.~L.~Wang, L.~Y.~Xiao, X.~H.~Zhong and Q.~Zhao,
  Understanding the newly observed $\Omega_c$ states through their decays,
  Phys.\ Rev.\ D {\bf 95}, no. 11, 116010 (2017).

\bibitem{Chen:2017sci}
  H.~X.~Chen, Q.~Mao, W.~Chen, A.~Hosaka, X.~Liu and S.~L.~Zhu,
  Decay properties of $P$-wave charmed baryons from light-cone QCD sum rules,
  Phys.\ Rev.\ D {\bf 95}, no. 9, 094008 (2017).

\bibitem{Agaev:2017lip}
  S.~S.~Agaev, K.~Azizi and H.~Sundu,
  Interpretation of the new $\Omega_c^{0}$ states via their mass and width,
  Eur.\ Phys.\ J.\ C {\bf 77}, no. 6, 395 (2017).

\bibitem{Wang:2017vnc}
  W.~Wang and R.~L.~Zhu,
  Interpretation of the newly observed $\Omega_c^0$ resonances,
  Phys.\ Rev.\ D {\bf 96}, no. 1, 014024 (2017).

\bibitem{Padmanath:2017lng}
  M.~Padmanath and N.~Mathur,
  Quantum Numbers of Recently Discovered $\Omega^{0}_{c}$ Baryons from Lattice QCD,
  Phys.\ Rev.\ Lett.\  {\bf 119}, no. 4, 042001 (2017).

\end{thebibliography}
\end{document}